\documentclass{article}
\usepackage{spconf,amsmath,epsfig}

\usepackage{booktabs}
\usepackage{url}
\usepackage{xcolor}
\usepackage{graphicx}
\title{Soil nitrogen forecasting from environmental variables provided by multisensor remote sensing images}
\threeauthors
  {Weiying Zhao, Ganzorig Chuluunbat, Aleksei Unagaev}
  {Deep Planet \\London, UK }
  {Natalia Efremova}
  {Queen Mary University and \\the Alan Turing Institute,	London, UK}
    {}
  {\\	}
\begin{document}
%\ninept
%
\maketitle
\begin{abstract}

%Original abstract. This study introduces a framework for forecasting soil nitrogen content, leveraging multisensor remote sensing images and advanced machine learning methods. Utilizing the Land Use/Land Cover Area Frame Survey (LUCAS) database, which spans Europe, we integrate various environmental variables from satellite sensors, distinct from previous studies, employ an extensive set of features and a broad spectrum of machine learning algorithms, focusing on tree-based models such as CatBoost, LightGBM, and XGBoost. The analysis covers diverse land covers like croplands and grasslands, ensuring the wide applicability of our findings. The CatBoost model exhibits exceptional performance in soil nitrogen estimation, surpassing other methods in accuracy. This research not only advances the field of agricultural management and environmental monitoring but also demonstrates the significant potential of integrating multisensor remote sensing data with machine learning for environmental analysis.

This study introduces a framework for forecasting soil nitrogen content, leveraging multi-modal data, including multi-sensor remote sensing images and advanced machine learning methods. We integrate the Land Use/Land Cover Area Frame Survey (LUCAS) database, which covers European and UK territory, with environmental variables from satellite sensors to create a dataset of novel features. We further test a broad range of machine learning algorithms, focusing on tree-based models such as CatBoost, LightGBM, and XGBoost.  We test the proposed methods with a variety of land cover classes, including croplands and grasslands to ensure the robustness of this approach. Our results demonstrate that CatBoost model surpasses other methods in accuracy. This research not only advances the field of agricultural management and environmental monitoring but also demonstrates the significant potential of integrating multisensor remote sensing data with machine learning for environmental analysis.

% The abstract should appear at the top of the left-hand column of text, about
% 0.5 inch (12 mm) below the title area and no more than 3.125 inches (80 mm) in
% length.  Leave a 0.5 inch (12 mm) space between the abstract's end and the main text's beginning.  The abstract should contain about 100 to 150
% words, and should be identical to the abstract text submitted electronically
% along with the paper cover sheet.  All manuscripts must be in English, printed
% in black ink.
\end{abstract}
\section{Introduction}
% Differences: test with LUCAS database which covers most of the European area.
% Tested several new features. 
\label{sec:intro}

Soil nitrogen is a crucial nutrient in agricultural ecosystems, influencing plant growth, soil fertility, and environmental sustainability. Accurate forecasting of soil nitrogen content is important for optimizing agricultural practices, enhancing crop yields, and minimizing environmental impacts such as nitrogen leaching and greenhouse gas emissions \cite{diaz2022machine}.

In recent years, the integration of remote sensing technology has revolutionized the field of agricultural monitoring and environmental management \cite{tuia2023artificial}. Multi-sensor remote sensing, involving data from various satellite sensors, offers a comprehensive approach to soil analysis. These sensors, which include optical, thermal, and microwave instruments, provide a broad spectrum of features that can be combines to give a more complete picture of soil conditions. For instance, optical sensors like Landsat capture high-resolution images that are useful for assessing vegetation cover and land use patterns, indirectly indicating soil nitrogen levels. Microwave sensors, such as those on Sentinel-1, effectively capture soil moisture content, a parameter closely related to soil nitrogen availability \cite{zhang2023comparison}. Sentinel-2, with its multispectral capabilities, further enhances this analysis by providing detailed information on vegetation health and soil characteristics \cite{zhou2023national}.

Integrating data from these diverse sources requires advanced data processing and analysis techniques. Machine learning (ML) algorithms, particularly those tailored for tabular data \cite{mcelfresh2023neural}, have shown significant promise in this domain. By efficiently handling large volumes of multi-sensor remote sensing data, ML algorithms can can capture complex patterns and relationships between observed environmental variables and soil nitrogen levels.

In this paper, we explore the application of machine learning methods, particularly focusing on tree-based algorithms, to forecast soil nitrogen content using multi-sensor remote sensing images.  Our methodology leverages the complementary strengths of different satellite sensors, coupled with robust machine learning techniques, to provide accurate and spatially detailed predictions of soil nitrogen across varied landscapes. The proposed approach differs from previous work \cite{ballabio2019mapping}, in the number of features and ML methods. Compared with \cite{patriche2023spatial}, the LUCAS database \cite{d2020harmonised}  used in this paper, has much larger spatial coverage, which requires different approach for feature preparation. This novel approach offers valuable insights for agricultural management, environmental monitoring, and policy-making in sustainable land use.

\section{Methodology}

\begin{figure*}
  \centering
  \includegraphics[width=0.8\textwidth]{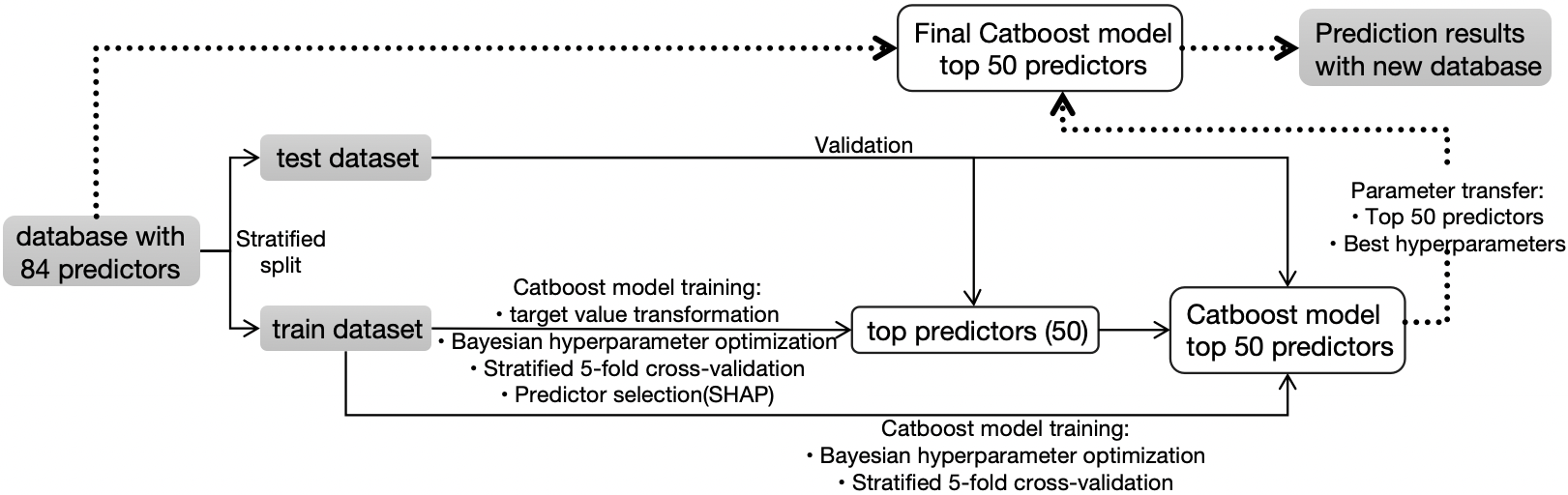}
  \caption{Example of the machine learning workflow for soil nitrogen estimation at 30 m resolution.}
   % Taking the CatBoost algorithm as a representative example.
  \label{fig:method_flowchart}
\end{figure*}

The workflow demonstrates the systematic approach we adopted from data preparation to model deployment (Fig.\ref{fig:method_flowchart}). 
Before model training, we implemented a normalization process to address the skewed distribution of soil nitrogen values. This involved scaling the target values by a factor of 100, followed by a logarithmic transformation to approximate a Gaussian distribution to enhance the predictive performance and stability of the model. We did a stratified dataset split based on landcover type, allocating 85\% for training and 15\% for testing for each landcover.  We performed feature selection using SHAP values, taking the top 50 features to reduce model complexity and the risk of overfitting, and enhancing interpretability and computational efficiency.

We performed hyperparameter tuning of the Catboost regressor via Bayesian optimization, coupled with 5-fold cross-validation. The objective was to optimize the model with respect to the Root Mean Square Error (RMSE) metric. This approach recursively refines the hyperparameter space based on previous evaluations, speeding up the tuning process. The training set was further stratified within each cross-validation fold to align with the target variable distribution. 
We chose RMSE as the performance metric for its robustness, ease of interpretation, and emphasis on larger prediction errors, which is crucial for model reliability.

After hyperparameter optimization and feature selection, we trained the final Catboost model on the entire dataset. The resulting model can be used on novel, previously unseen datasets due to its generalizability and the effectiveness of our methodological approach.

\section{Experimental results and discussion}
% \textcolor{red}{Database introductions one by one, reason, usefulness}

\subsection{Dataset}

% \begin{figure}[h]
%   \centering
%   \includegraphics[width=0.5\textwidth]{results/database_measure_time_visualization.pdf}
%   \caption{Distribution of database measuring time. The times have been converted to the days of the year.}
%   \label{fig:database_measure_time_visualization}
% \end{figure}

We compared four tree-based methods using the LUCAS database \cite{d2020harmonised}, a harmonized dataset encompassing in-situ land cover and land use across the European Union. All methods shared a comparison framework.
We used the following remote sensing sources to prepare a multi-modal dataset \cite{zhao2023soil}:
\begin{itemize}
\itemsep0em  % Set spacing to 0 for this list only
    \item COPERNICUS/USGS DEMs: terrain attributes, such as the slopes, aspects, elevations
    \item ERA5-Land Daily Aggregated data (ECMWF): annual statistical features of soil temperature, runoff, surface net solar radiation, evaporation, precipitation, etc.
    \item Landsat-8: monthly median of cloud-free multispectral band values around the soil nitrogen measured time and related vegetation indices
    \item MODIS: the annual statistics, percentile or cumulative values of GDD, temperature, PET, FPAR, LAI, etc.
    \item Sentinel-1: monthly mean of single polarization values around the soil nitrogen measurement time and radar vegetation indices which provide soil surface reflecting signals
    \item CHIRPS Daily: annual rainfall statistics
    \item Measurement attributes: season, days of year, geographical location
    \item OpenLandMap: sand, clay and texture
\end{itemize}

% The terrain attributes are derived from COPERNICUS and USGS DEMs. Macroclimate features come from ERA5-Land Daily Aggregated data (ECMWF). Landsat-8 offers high-resolution landcover images, MODIS supplies medium-resolution surroundings, and OpenLandMap contributes some soil attributes, Sentinel-1 data provide surface reflecting signals, CHIRPS Daily provides the annual rainfall statistics and also some measuring features, such as day of year and season, etc. 

All the data was prepared with the help of Google Earth Engine.\footnote{\url{https://earthengine.google.com/}}
The resulting tabular format dataset consists of 21244 samples with 84 features each, which were collected from croplands (13937) and grasslands (7307) in 2015 and 2018. 

%The data split was 70\% training, 15\% testing, and 15\% evaluation.

% Sloping landscapes dominate the Earth's land surface. Every year, soil erosion laterally distributes 75 Gt of topsoil (Berhe et al. 2007). The coupled biogeochemical cycles of carbon (C) and nitrogen (N) are strongly influenced by soil erosion as it affects their fluxes in and out of the soil system, storage, distribution within the soil matrix, and residence time in soil. At the global scale, it is estimated that erosion can account for a net sink for atmospheric carbon dioxide.

% \subsection{Experimental setup}

\begin{table*}[ht]
  \caption{Soil nitrogen estimation accuracy based on the test dataset. Best MAPE and MAE are in bold.}
  \label{tab:spatial_interpolation}
  \centering
  \small
  \begin{tabular}{lllllllll}
    \toprule
Method&	MAPE\_total&	MAPE\_crop&	MAPE\_grass&	MAE\_total&	MAE\_crop&	MAE\_grass&	Features&	Parameters\\
\midrule
CatBoost&	\textbf{29.019}&\textbf{25.506}&35.722&\textbf{0.609}&	\textbf{0.391}&\textbf{1.026}&all&default\\
LightGBM&	29.779&	26.527&	35.981&	0.622&	0.405&	1.035&	all&	default\\
XGBoost&	30.310&	26.625&	37.340&	0.636&	0.413&	1.062&	all&	default\\
ExtraTrees&	29.262&	27.250&	\textbf{33.100}&	0.632&	0.403&	1.068&	all&	default\\
\midrule
CatBoost&	\textbf{28.742}&	\textbf{25.364}&	35.187&	\textbf{0.605}&	\textbf{0.390}&	\textbf{1.014}&	selected&	default\\
LightGBM&	29.543&	26.217&	35.89&	0.617&	0.401&	1.030&	selected&	default\\
XGBoost&	30.489&	26.818&	37.491&	0.637&	0.416&	1.058&	selected&	default\\
ExtraTrees&	29.097	&26.977&	\textbf{33.143}&	0.629&	0.401&	1.064&	selected&	default\\
\midrule								
CatBoost&	28.466&	\textbf{24.906}&35.257&	0.602&	\textbf{0.386}&	1.016&	selected&	optimized\\
LightGBM&	\textbf{28.463}&	25.153&	34.778&	\textbf{0.599}&	0.387&	\textbf{1.004}&	selected&	optimized\\
XGBoost&	28.903&	25.336&	35.707&	0.608&	0.39&	1.026&	selected&	optimized\\
ExtraTrees&	28.866&	26.757&	\textbf{32.891}&	0.623&	0.396&	1.057&	selected&	optimized\\
  \bottomrule
  \end{tabular}
\end{table*}

\subsection{Results and discussion}
Upon evaluating soil nitrogen estimation accuracy across four distinct methodologies on the test dataset (Tab.\ref{tab:spatial_interpolation}), the CatBoost model showed the best performance in most feature and hyperparameter settings. Notably, when using all features with default parameters, CatBoost achieved the lowest MAPE\_total of 29.019 and MAE\_crop of 0.391, indicating a more precise estimation for both overall and crop-specific nitrogen levels. In contrast, LightGBM demonstrated marginally higher error rates.

A comparison of models with selected features and default parameters revealed consistent improvements, with CatBoost again leading with a reduced MAPE\_total of 28.742 and MAE\_crop of 0.390. Similarly, XGBoost and ExtraTrees demonstrated enhanced accuracy with selected features, albeit still trailing behind CatBoost.
Further feature selection and parameter optimization showed significant improvement in accuracy for LightGBM, achieving the best MAPE\_total of 28.463 and the lowest MAE\_grass of 0.599, demonstrating the importance of hyperparameter optimization. CatBoost provided better results on MAPE\_crop and MAE\_crop.
XGBoost and ExtraTrees also showed improved results with optimized parameters, though they did not surpass LightGBM's and CatBoost's performance.
These findings underscore the critical impact of feature selection and hyperparameter optimization on model accuracy.
 
% \begin{figure}[htb]
\begin{figure*}
  \centering
  \vspace{-20pt} % Adjust the space before the figure
  \includegraphics[width=1\textwidth]{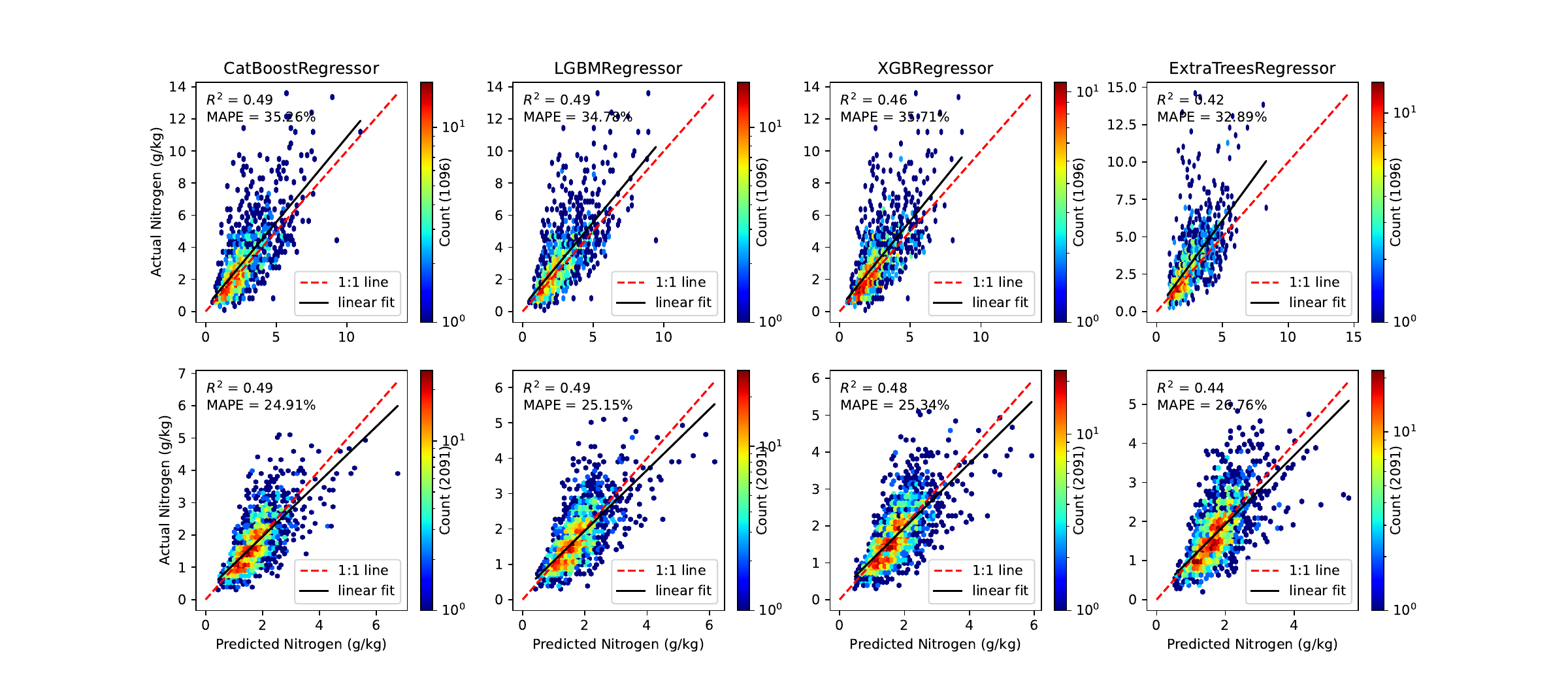}
  \caption{Scatter plot comparison of observed versus predicted soil nitrogen values provided by various methodologies using the test database. Top: soil nitrogen measured in grassland, bottom: cropland.
  The methods are trained according to the framework introduced in Fig.\ref{fig:method_flowchart} with selected top 50 features and optimized hyperparameters.}
  \label{fig:scatter_plot_eval_part}
\end{figure*}

Fig.\ref{fig:scatter_plot_eval_part} shows moderate correlation between actual and predicted nitrogen levels for all four models, as indicated by the $R^2$ values around 0.5. The density of points along the 1:1 line in each plot shows the predictive accuracy across the range of nitrogen values. The closer the points cluster to this line, the more accurate the predictions are. The spread of points away from the line, especially in the ExtraTreesRegressor, indicates a tendency towards over or under-prediction in certain nitrogen value ranges. CatBoostRegressor and LGBMRegressor both demonstrate an $R^2$ of 0.49, suggesting nearly identical performance in terms of variance explanation, with CatBoostRegressor having a slightly lower MAPE at 24.91\% compared to LGBM's 25.15\% over cropland.
XGBRegressor's performance is marginally lower with an R² of 0.48 and a MAPE of 25.34\%, indicating a somewhat less accurate prediction capability. ExtraTreesRegressor shows a more pronounced decrease in performance with the lowest R² of 0.44 and the highest MAPE at 26.76\%, suggesting that it may be less effective at capturing the nuances of soil nitrogen levels in this specific test database.

\begin{table}
  \caption{Soil nitrogen estimation accuracy in cropland: Test MAE and MAPE scores using four tree-based models. Data split 85/15.}
  \label{tab:spatial_interpolation_crop}
  \centering
  \small
  \begin{tabular}{lllll}
    \toprule
Method&	MAPE& MAE&	Features&	Parameters\\
\midrule
CatBoost&	\textbf{23.356}&	\textbf{0.370}&	all features&	default\\
LightGBM&	24.104&	0.382&	all features&	default\\
XGBoost&	24.800&	0.390&	all features&	default\\
ExtraTrees&	24.317&	0.379&	all features&	default\\
\midrule
CatBoost&	\textbf{23.077}&	\textbf{0.367}&	selected features&	default\\
LightGBM&	24.070&	0.381&	selected features&	default\\
XGBoost&	24.760&	0.388&	selected features&	default\\
ExtraTrees&	23.928&	0.373&	selected features&	default\\
  \bottomrule
  \end{tabular}
\end{table}

% \begin{figure}[htb]
\begin{figure*}[ht]
  \centering
  \includegraphics[width=0.88\textwidth]{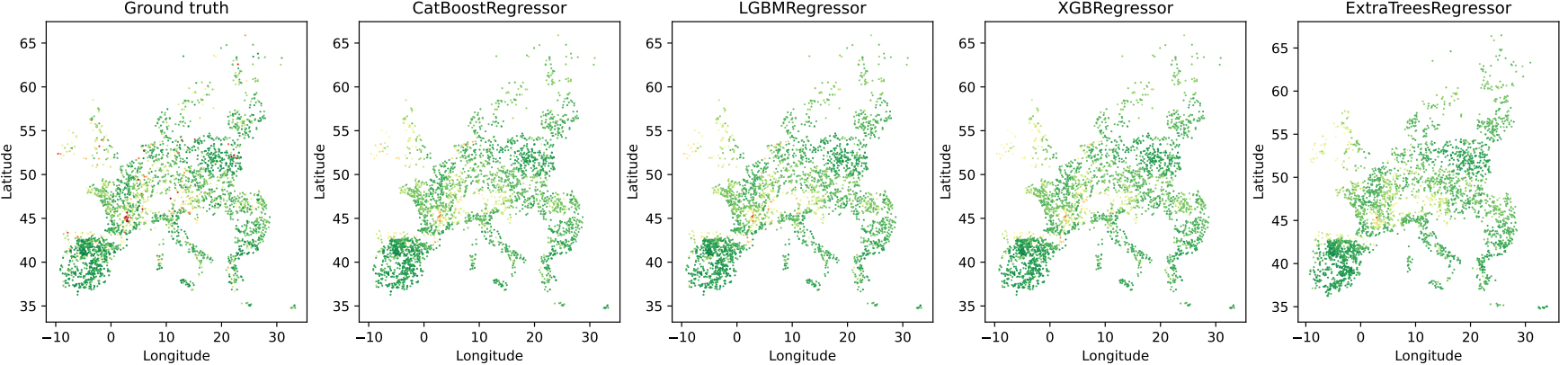}
  \caption{Scatter plot of real soil nitrogen and predicted soil nitrogen provided by different methods on the test database.}
  \label{fig:prediction_results_comparison}
  \vspace{-10pt} % Adjust the space after the figure
\end{figure*}

If the target value range differs from other landcover types, the model may provide better results if only trained with the database measured in the same landcover type. As shown in Tab.\ref{tab:spatial_interpolation_crop}, the MAPE and MAE provided by different methods are better than those shown in Tab.\ref{tab:spatial_interpolation}.

While all models present a comparable degree of correlation between predicted and actual values, their prediction results show similar spatial variance as shown in Fig.\ref{fig:prediction_results_comparison}. CatBoostRegressor slightly outperforms the others under different hyperparameter and feature settings, with LGBMRegressor following closely. The ExtraTreesRegressor, while still within a competitive range, shows room for improvement, particularly in model precision and consistency across the range of predictions. Due to the multiscale nature of the used predictors, the worst prediction cases are mainly located in areas with heterogeneous landcovers, like near a tree, river, building, etc.

\section{Conclusions}

% Apply the framework to different Different landcovers.
% Provide a reference for the study and use of LUCAS database. 

This study provided a framework to forecast soil nitrogen content using multisensor remote sensing images and machine learning, focusing on tree-based models like CatBoost, LightGBM, and XGBoost. All models \cite{mcelfresh2023neural} good at handling tabular datasets are applicable to this task.
Using the extensive LUCAS database, our methodology integrates diverse environmental variables from satellite sensors, offering a comprehensive approach to soil analysis. The CatBoost model demonstrated exceptional performance, particularly in terms of accuracy in soil nitrogen estimation, highlighting the potential of combining remote sensing data with machine learning in environmental analysis.
The findings demonstrate the effectiveness of our methodological approach in handling large volumes of data and discerning complex patterns. The study's emphasis on diverse land covers like croplands and grasslands ensures wide applicability and relevance to various ecosystems. Future research will include the selection of a training dataset concerning the feature resolutions and land classification maps.

% References should be produced using the bibtex program from suitable
% BiBTeX files (here: strings, refs, manuals). The IEEEbib.bst bibliography
% style file from IEEE produces unsorted bibliography list.
% -------------------------------------------------------------------------
\small
\bibliographystyle{IEEEbib}
\bibliography{refs}
% \bibliography{strings,refs}

\end{document}